\documentclass[epj]{webofc}
\usepackage[varg]{txfonts}   
\usepackage{hyperref}
\newcommand*{\tabref}[1]{Table~\ref{tbl:#1}}
\newcommand*{\tablab}[1]{\label{tbl:#1}}
\renewcommand*{\eqref}[1]{Eq.~(\ref{eq:#1})}

\newcommand*{\figref}[1]{Fig.~(\ref{fig:#1})}
\newcommand*{\figlab}[1]{\label{fig:#1}}

\newcommand*{\seclab}[1]{\label{sec:#1}}
\usepackage{upgreek}
\usepackage[tight]{subfigure}

\newcommand{\comment}[1]{}

\wocname{ARENA-2016}
\woctitle{ARENA-2016}
\begin{document}
\title{Acoustic detection of high energy neutrinos in sea water: status and prospects}
%
%

\author{\firstname{Robert} \lastname{Lahmann}\inst{1}\fnsep\thanks{\email{robert.lahmann@fau.de}}
}


\institute{Friedrich-Alexander-Universit\"at Erlangen-N\"urnberg, Erlangen Centre for Astroparticle Physics (ECAP), Erwin-Rommel-Stra\ss e 1, 91058 Erlangen, Germany
          }

\abstract{%
The acoustic neutrino detection technique is a promising approach for future large-scale detectors with the aim of measuring the small expected flux of 
neutrinos at energies in the EeV-range and above. The technique is based on the thermo-acoustic model, which implies that the energy deposition by a particle cascade -- resulting from a neutrino interaction in a medium with suitable thermal and acoustic properties -- leads to a local heating and a subsequent characteristic pressure pulse that propagates in the surrounding medium. 
%
Current or recent test setups for acoustic neutrino detection have either been add-ons to optical neutrino telescopes or have been using acoustic arrays built for other purposes, typically for military use. While these arrays have been too small to derive competitive limits on neutrino fluxes, they allowed for detailed studies of the experimental technique. 
%
With the advent of the research infrastructure KM3NeT in the Mediterranean Sea, new possibilities will arise for acoustic neutrino detection.  
In this article, results from the ``first generation'' of acoustic arrays will be summarized and 
implications for the future of acoustic neutrino detection will be discussed.
}
\maketitle
\section{Introduction}
\label{intro}
\seclab{intro}
\comment{ 
``Of all the forms of radiation known, sound  travels through the sea the 
best''~\cite{urick}.
Hence, sound is 
used in the sea by marine mammals and by humans for purposes of 
communication and positioning.
It is important for astroparticle physics that 
sound waves are furthermore emitted when the local medium heats up 
following the interaction of a neutrino in water. 
As will be elaborated in Sec.~\ref{sec:acou_det_nus}, this effect
allows for the detection of ultra-high-energy neutrinos.
Apart from the design of the detector, the energy threshold is essentially
determined by the
relatively high ambient noise in the sea and by the small
signals expected from neutrino interactions.
} 


Measuring acoustic pressure pulses in huge underwater acoustic arrays
is a promising approach for the detection of 
ultra-high-energy (UHE, $E_{\nu} \gtrsim 10^9$\,GeV) neutrinos.
These
are expected to be produced in interactions of cosmic rays with the cosmic microwave background~\cite{bib:Berezinsky-1969}.
The pressure signals are produced by the
particle showers that evolve when neutrinos interact with nuclei in
water.
The resulting energy deposition of the hadronic cascade within a cylindrical volume of a few
centimetres in radius and several metres in length leads to a local
heating of the medium, see e.g.~\cite{bib:Erlangen_testbeam-2015} and references therein.

According to the so-called thermo-acoustic model
\cite{Askariyan2}, the energy deposition of particles traversing
liquids leads to a local heating of the medium which can be regarded
as instantaneous with respect to the hydrodynamic time scales. Due to
its temperature change, the medium expands or contracts according to
its bulk volume expansion coefficient
. The accelerated motion of
the heated medium generates a pulse whose temporal signature
is bipolar with a spectrum peaking around 10\,kHz,  propagating through the medium. 
Coherent superposition of the elementary sound waves, produced over the cylindrical volume of the energy deposition, 
leads to a propagation within a flat disk-like volume, often referred to as ``pancake'', in the direction perpendicular to the axis of the particle shower.
 The main advantage of using sound for the detection of neutrino interactions, as opposed to Cherenkov light, lies in the much longer attenuation length of the former type of radiation -- several kilometres for sound compared to several tens of meters for light in the respective frequency ranges of interest in sea water.

In addition to water and ice, 
which is the medium of acoustic detection test experiments 
presently or recently conducted,
acoustic detection in salt domes\ \cite{bib:salt-acoustics-arena2005,bib:price-2006} and in 
permafrost\ \cite{bib-permafrost-2008}
has been discussed.
%
In this article, the detection of neutrinos in sea water will be discussed, with an emphasis on the achievements so far and the lessons learned for the future.
Discussions on neutrino detection in fresh water and ice can be found in \cite{bib:budnev-arena2016} and \cite{bib:karg-arena2012}, respectively.
%
%

\section{Sound in water and acoustic neutrino detection}

Two processes must be understood to asses the potential of acoustic neutrino detection:
The conversion of the energy released by a neutrino interaction into a detectable acoustic signal and the attenuation of the resulting sound wave in the medium. 

\paragraph{Acoustic signal:}
For a simplified analytic derivation it can be assumed that the shower energy in the plane perpendicular to the 
direction of the incoming neutrino 
is described by a Gaussian distribution 
with standard deviation $\sigma_\rho$. If the total energy deposited in the hadronic shower is denoted
by $E_0$, then $p_\text{max}$, i.e.\ half the peak-to-peak amplitude of the bipolar pulse -- measured as a function of time at a fixed distance in the far field 
-- is given by
\begin{equation}
p_\text{max} \propto \gamma_\text{G} \frac{E_0}{\sigma_\rho^2}
\end{equation}
where the dimensionless quantity 
$
\gamma_G \equiv v_s^2 \alpha/c_p
$ 
is the Gr\"uneisen parameter, which depends on the speed of sound $v_s$, the bulk volume expansion coefficient
$\alpha$ and the specific heat capacity at constant pressure $c_p$. For a derivation, see~\cite{Learned} or e.g.~\cite{bib:lahmann-habil-2011} and references therein.
In \figref{fig:gruneisen}, the Grüneisen parameter is shown for several bodies of water. For the Mediterranean Sea, its value is quite favourable for acoustic neutrino detection, basically due to its relatively high temperature of $\sim$%
$13^\circ$C even at great depth. 
For Lake Baikal, while otherwise well suited for acoustic neutrino detection, the Grüneisen parameter is small since in the deep zone of the lake, the water temperature is only 
$1.5-2^\circ\mathrm{C}$ higher than
the maximum density at the respective 
depth~\cite{bib:baikal-arena2010}. 
Reflected by the Grüneisen parameter is the temperature change vs.\ depth in the top layers of the respective bodies of water. For greater depth, where the temperature gradient becomes small, the Grüneisen parameter basically follows
the profile of the speed of sound, which increases almost linearly with depth due to the increasing pressure. 
The temperature profile for the Mediterranean Sea and
tropical ocean waters is shown in \figref{fig-2}. 

\begin{figure*}[h]
\centering
\includegraphics[width=7.1cm,clip]{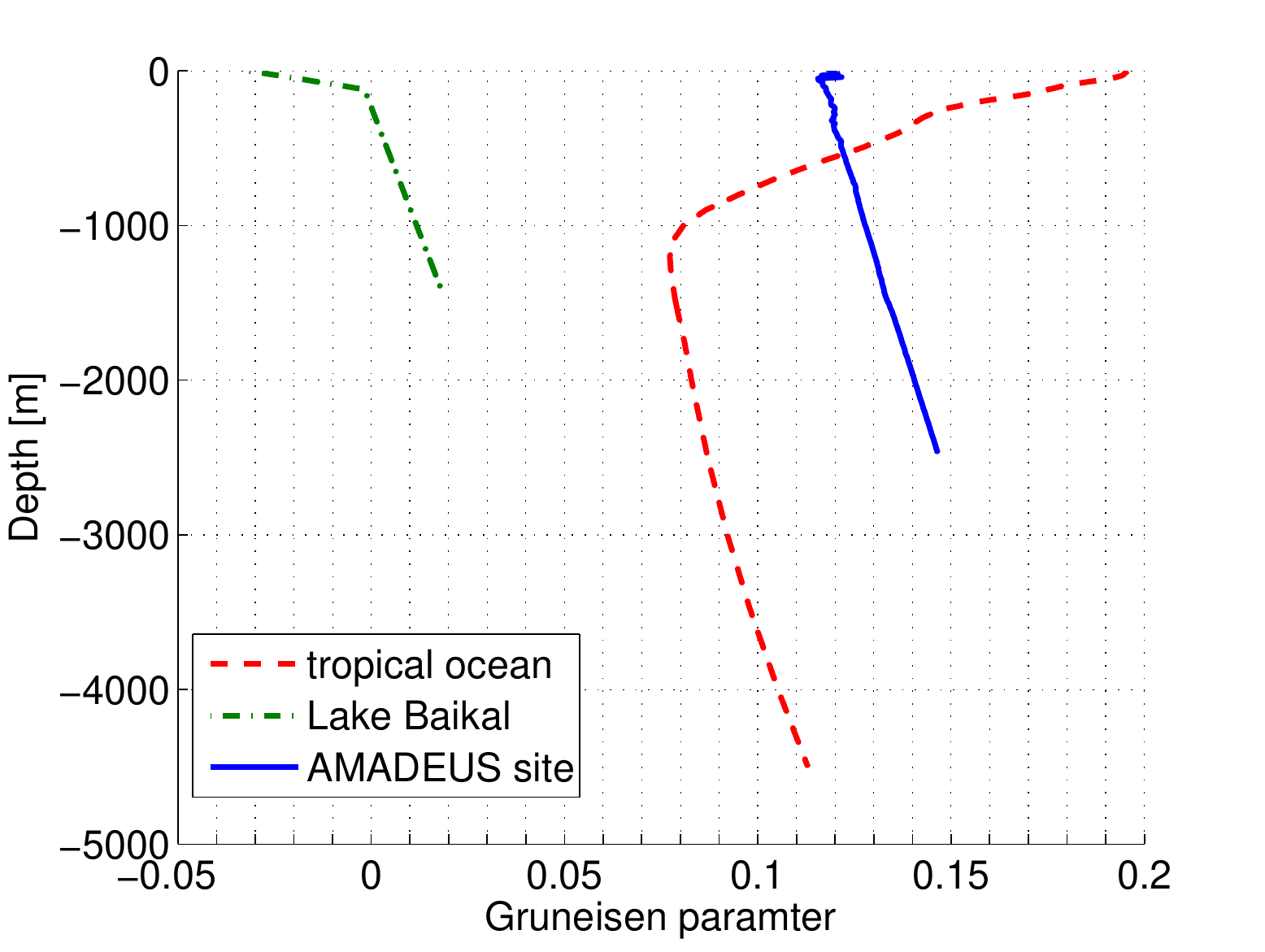} 
\caption{
Grüneisen parameter for Lake Baikal (fresh water), Mediterranean Sea at the ANTARES site, and tropical ocean waters (sea water).
 Figure from~\cite{bib:lahmann-habil-2011}. 
}
\figlab{fig:gruneisen}       
\end{figure*}

\begin{figure*}[th]
\centering
\includegraphics[width=7.0cm,clip]{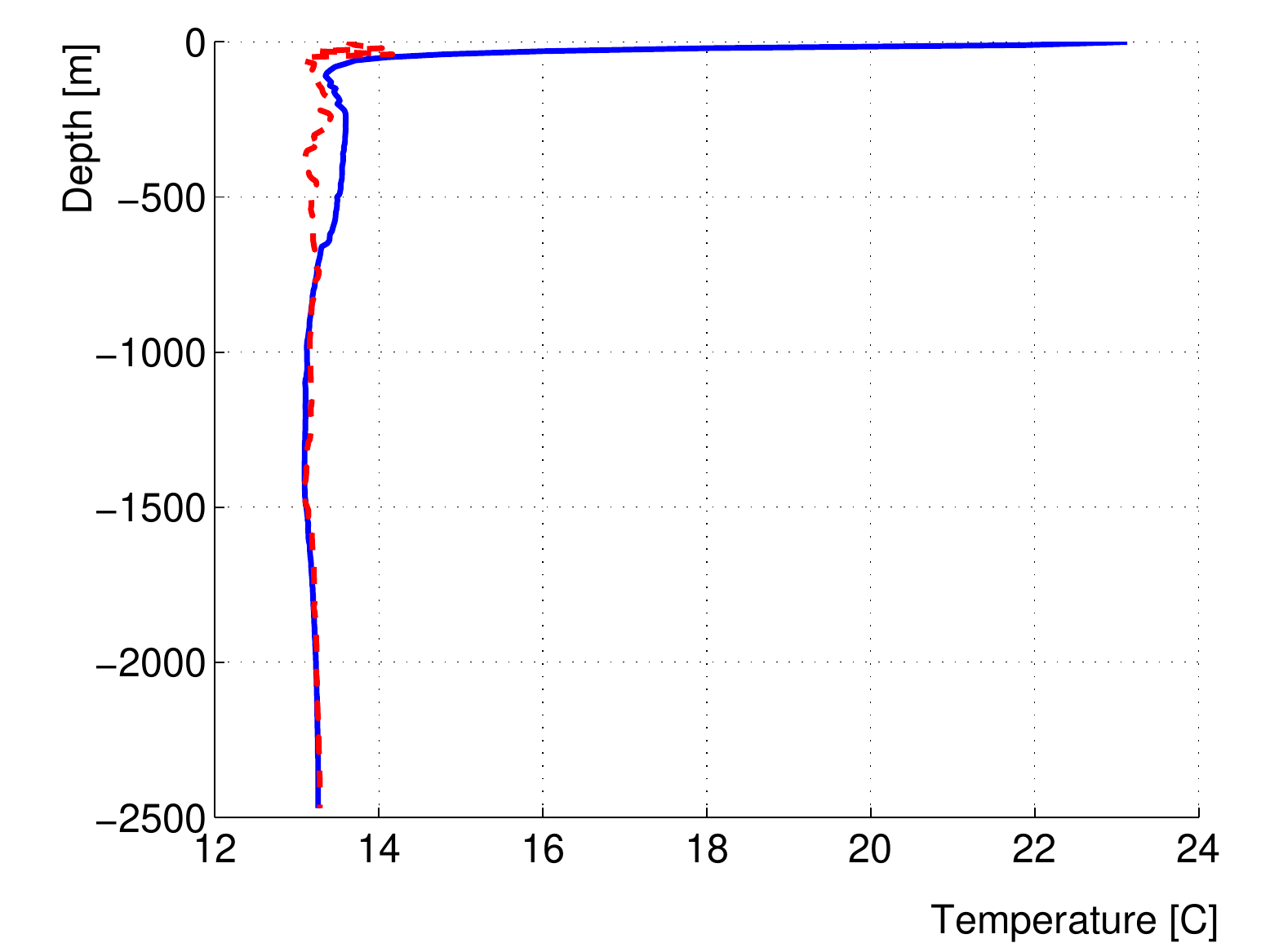}
\hspace*{-5mm}
\includegraphics[width=7.0cm,clip]{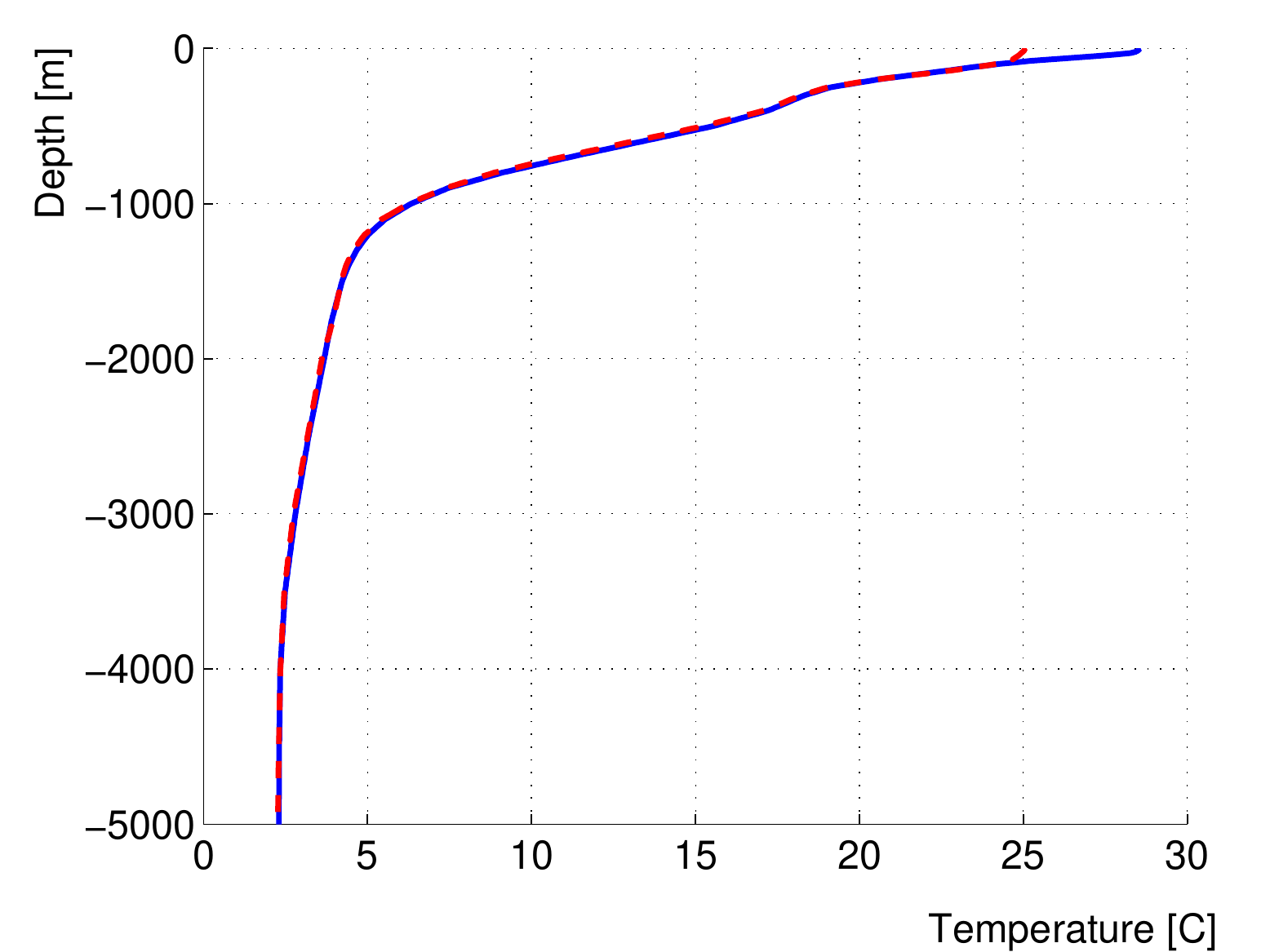}
\caption{
Temperature as a function of the depth below the sea surface, for
different seasons at two different sites;
\textit{left}: Measurement from the ANTARES site in
August 2007 
(solid blue line) and in March
2010 
(dashed red line);
\textit{right}:
Temperature profile in the tropical
ocean
, 24$^\circ$30'N and 72$^\circ$30'W, for the
average of the three summer months (solid blue line) and the three
winter months (dashed red line).  
Figure from~\cite{bib:lahmann-habil-2011}.
}
\figlab{fig-2}       
\end{figure*}

\paragraph{Attenuation:}
Attenuation in distilled water is caused by viscous absorption and is proportional to the square of the
frequency of the sound wave.  In
sea water, in addition ionic relaxation of chemical compounds
solved in the water contributes.
The latter effect is due to the dependence of the association
$\rightleftharpoons$ dissociation process of the chemical compounds on
temperature and pressure~\cite{bib:Liebermann-1949}.
%

The frequency-dependent attenuation of sound in water is shown in \figref{fig:absorption}.
Although NaCl is the principal constituent of salt in sea water and 
magnesium sulphate MgSO$_4$  only amounts for a fraction
of 4.7\,\% by weight of the total dissolved salts in sea water, 
the latter dominates the absorption process
for frequencies above a few kHz, up to about 100\,kHz~\cite{urick}, i.e.\
in the region relevant for acoustic detection.
The absorption length at the ANTARES site is about 5\,km at 10\,kHz, compared to about 100\,km
for distilled water. 
Clearly, if conditions provide for Grüneisen parameters of equivalent size, fresh water is superior to sea water for acoustic neutrino detection as it allows to survey a  larger volume with the same number of sensors\footnote{%
The actual layout and instrumentation density of an acoustic neutrino telescope also depends on further factors, such as the noise background and the ``pancake'' shape of the three-dimensional pressure field.
}.

\begin{figure*}[h]
\centering
\includegraphics[width=8.2cm]{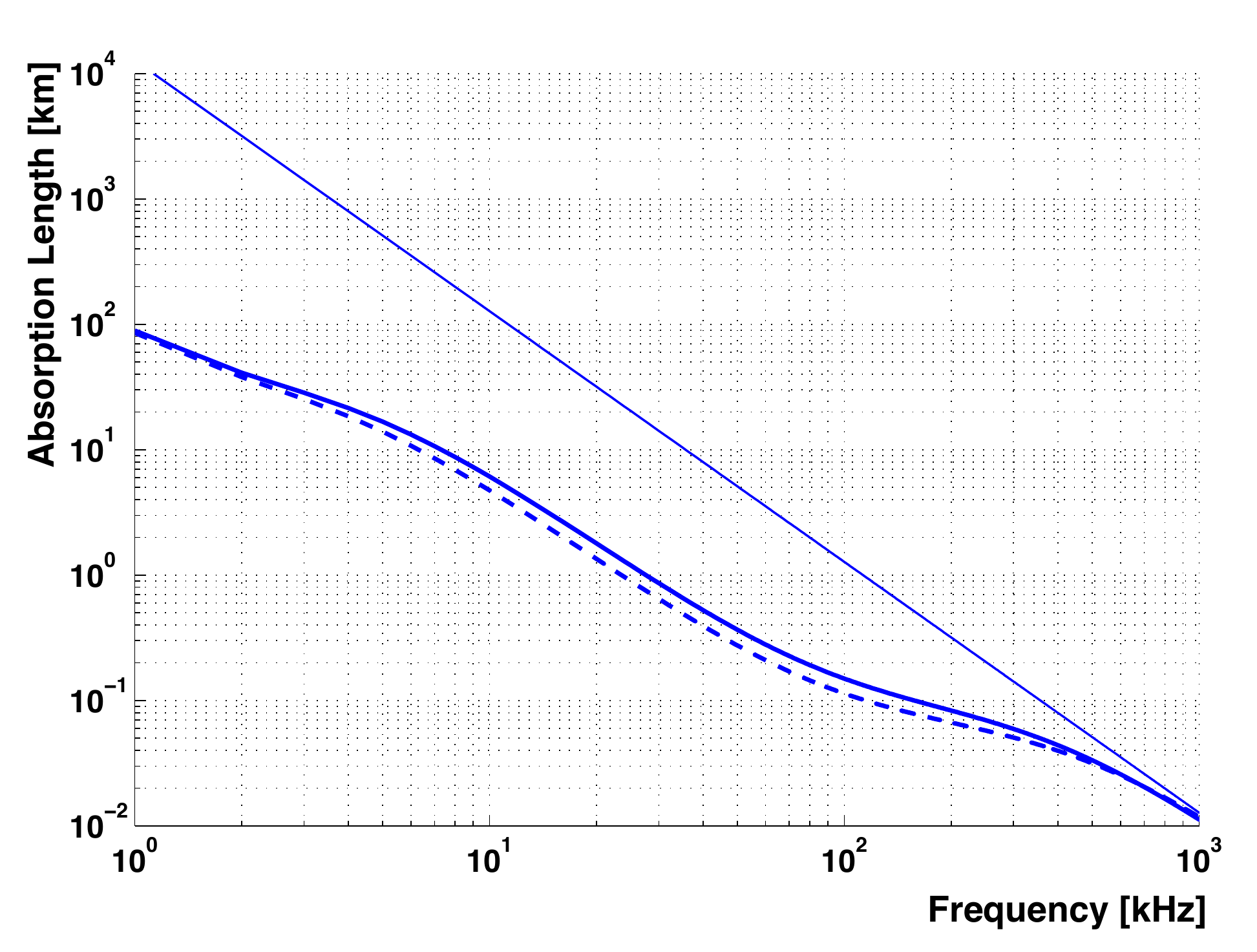}
\caption{Absorption length according to the parametrisation given 
in~\cite{bib:Ainslie-98} for conditions at the ANTARES site 
with temperature $T = 13.2^\circ$\,C, salinity of 3.8\,\% and a pH-value of 8. 
The solid thick curve is for a depth of 2000\,m, the broken thick curve
for a depth of 100\,m. The straight thin line is the absorption length for 
pure water at a depth of 2000\,m.  Figure from~\cite{bib:lahmann-habil-2011}.
}
\figlab{fig:absorption}
\end{figure*}

\section{Results of the ``first generation'' of acoustic neutrino test setups}

Current or recent test setups for acoustic neutrino detection have either been add-ons to 
optical neutrino telescopes or have been using acoustic arrays built for other purposes, 
typically for military use. The acoustic sensors are based on the piezoelectric effect~\cite{anton} and $\cal{O}$(10) of these sensors have been installed in the respective setups.
%
%
An overview of these ``first generation'' experiments in sea water -- and for completeness also in fresh water and ice -- that are currently taking data or have done
so until recently is given in \tabref{tab:acoustic_sites}.
%
Details on the experiments can be found in the references given in the table. 
At the writing of this article, only SPATS and the Lake Baikal experiment were still taking 
data\footnote{For the military arrays, this means that data are not used for investigations of 
neutrino detection methods any more}.

Limits on the flux of UHE neutrinos have been derived by the SPATS~\cite{bib:spats-2011-bkgr}, SAUND~\cite{bib:saund2010} and ACoRNE~\cite{bib:bevan_ARENA08} experiments.
 While the size of the experiments is too small to derive competitive limit, 
it could be shown that the tools and techniques for setting such limits for acoustic neutrino detection are in place.

\begin{table*}[hb]
\centering
\caption{Overview of existing and recent acoustic detection test sites.
}
\tablab{tab:acoustic_sites}      
\begin{tabular}{|l|l|l|l|l|}
\hline
 Experiment & Location & Medium & Sensor  & Host    \\
  &  &  &  Channels &  Experiment \\
\hline
\hline
 SPATS~\cite{bib:karg-arena2010,bib:SPATS-design-perform-2011}  & South Pole & Ice & 80 & IceCube
 \\ \hline
 Lake Baikal~\cite{bib:budnev-arena2016} & Lake Baikal  & Fresh Water & 4  & Baikal Neutrino 
  \\ 
  &  &  &   & Telescope 
   \\ \hline
 O$\upnu$DE~\cite{bib:noise_ONDE_arxiv} &  Mediterranean Sea  (Sicily)  & Sea Water & 4  &  NEMO\, 
 \\ \hline
 AMADEUS~\cite{bib:amadeus-2010}      &  Mediterranean Sea (Toulon)  & Sea Water & 36  & ANTARES\, 
 \\ \hline
 ACoRNE~\cite{bib:acorne}     &   North Sea (Scotland)  & Sea Water & 8  & Rona  military array    \\ \hline
 SAUND~\cite{bib:Lehtinen-2002}  &   Tongue of the Ocean   & Sea Water & 7/49$^{(\star)}$  & AUTEC military      \\ 
 & (Bahamas) & & & array \\
\hline
\hline
\multicolumn{5}{l}{$^{(\star)}${\footnotesize\,The number of hydrophones was increased from 7 in SAUND-I 
to 49 in SAUND-II}}
\end{tabular}
\end{table*}




To assess the potential for acoustic neutrino detection in a given natural body of water, 
ambient noise and transient background at the site 
have to be investigated.
%
As a next step beyond setups of the ``first generation'', the KM3NeT infrastructure, currently under construction in the Mediterranean Sea, will allow for investigations of acoustic neutrino detection in a volume of several cubic kilometres of sea water~\cite{bib:km3net_acou_arena2016}.
The KM3NeT neutrino telescope~\cite{bib:KM3NeT2p0_LOI_2016} will comprise a huge array of acoustic sensors for position calibration that can
be used for further studies of neutrino detection.
Concepts employing fibre-based hydrophones as an alternative to piezo-based sensors 
for a potential future extension to KM3NeT are discussed elsewhere~\cite{bib:buis_arena2016}.

Studies of the conditions in the Mediterranean Sea, as performed 
by the  O$\upnu$DE  and AMADEUS experiments, 
provide input for simulations for 
KM3NeT and beyond. 
The ambient noise is broadband and is mainly caused by agitation of the sea surface\,\cite{urick2}, i.e.~by rain, wind, breaking waves, spray, and cavitations. 
%
It is predominantly the ambient noise that determines the 
energy threshold for neutrino detection.
The ambient noise was measured to be low and stable by the O$\upnu$DE~\cite{bib:noise_ONDE_arxiv}  and AMADEUS~\cite{Kiessling_master-2013,bib:lahmann-amadeus-icrc2011} experiments, and was found to be generally favourable for the operation of 
an acoustic neutrino detector.

Transient noise signals have short duration and an amplitude exceeding the ambient noise level. When of bipolar shape, these signals can mimic pulses from neutrino interactions. Sources of such signals can be anthropogenic, e.g.\ shipping traffic, or
marine fauna. In particular dolphins emit short signals with a spectrum similar to that of acoustic emissions from neutrino interactions. 

%
The transient background, as measured at the ANTARES site, has a relatively high rate. 
It is essential to reduce this background completely in order to measure the low expected rate of UHE neutrinos. 
%
A strong background reduction is already achieved with methods developed with 
AMADEUS~\cite{bib:Neff_PhD_2013}.
Machine learning algorithms are used to identify bipolar pulses. Furthermore, events that are spatially and temporally clustered are discarded, as neutrinos are unlikely to produce several signals from one position within a short period of time. 
For further reduction it is essential to use the pancake shape of the signal as a selection
criterion. This is not possible with AMADEUS due to its small size -- such investigations can be pursued with the acoustic array of the KM3NeT detector~\cite{bib:kiessling_arena2016}.

In order to reduce the recorded data volume,
the AMADEUS system employs an online
pulse-shape-recognition trigger which is sensitive to the bipolar pulse expected from neutrino
interactions~\cite{bib:amadeus-2010}. This trigger selects events with a wide range of shapes. Improving the online selection may lead to a reduction of the signal-to-noise ratio of the data available for offline analysis, which would be an important step towards reducing the energy threshold for neutrino detection and thereby improving the sensitivity of a 
potential future acoustic neutrino telescope.


\section{Conclusions and outlook}
%
Acoustic arrays of the ``first generation'',
either existing military acoustic arrays or  additions to 
Cherenkov neutrino telescopes,
have been used to investigate neutrino detection methods. 
Their sizes are far too small to yield competitive limits on 
the flux of UHE neutrinos but they allow for the investigation of experimental techniques
for a future acoustic neutrino detector and for the investigation of background conditions
for feasibility studies.


In the Mediterranean Sea, the KM3NeT neutrino telescope is currently under construction.
It will comprise a 
system for acoustic position calibration of its optical sensors 
which
can also be used for  investigations towards acoustic neutrino
detection. This would be an intermediate step towards
an even bigger acoustic detector for UHE neutrinos. 
New concepts, in particular fibre-based hydrophones, can be used in the long run to further increase the instrumented volume of KM3NeT beyond the volume instrumented for
optical detection, 
see \cite{bib:buis_arena2016} and references therein.

\section{Acknowledgements}
The author wishes to thank the organisers of the ARENA workshop for
the invitation to give a presentation. The AMADE\-US project is supported
by the German government (Bundesministerium f\"ur Bildung und
Forschung, BMBF) through grants 05A08WE1 and 05A11WE1.

%
%
\bibliography{lahmann_general}

\begin{thebibliography}{32}

\bibitem{bib:Berezinsky-1969}
{V.S.~Berezinsky and G.T.~Zatsepin}, PL \textbf{B\,28}, 423 (1969)

\bibitem{bib:Erlangen_testbeam-2015}
{R.~Lahmann \etal}, Astropart.\ Phys. \textbf{65}, 69 (2015), arXiv:1501.01494
  [astro-ph.IM]

\bibitem{Askariyan2}
{G.A.\ Askariyan \etal}, Nucl.\ Inst.\ and Meth. \textbf{164}, 267 (1979)

\bibitem{bib:salt-acoustics-arena2005}
{G.~Manthei, J.~Eisenbl\"{a}tter, and T.~Spies}, Int.\ J.\ Mod.\ Phys.\
  \textbf{A21S1}, 30 (2006)

\bibitem{bib:price-2006}
{P.B. Price}, J.\ of Geophys.\ Res \textbf{111}, B02201 (2006),
  {arXiv:astro-ph/0506648v1}

\bibitem{bib-permafrost-2008}
{R.~Nahnhauer, A.A.~Rostovtsev, and D.~Tosi}, Nucl.\ Inst.\ and Meth.
  \textbf{A\,587}, 29 (2008)

\bibitem{bib:budnev-arena2016}
{N.\ Budnev}, \emph{{Acoustic detection of high energy neutrinos in fresh
  water}}, in \emph{{Proc.\ of ARENA 2016 workshop}} (2016), to be published in
  EPJ Web of Conf.

\bibitem{bib:karg-arena2012}
{T.~Karg}, {AIP Conf.\ Proc.\ } \textbf{1535}, 162 (2013), {arXiv:1210.7974
  [astro-ph.IM]}

\bibitem{Learned}
{J.G.~Learned}, Phys.\ Rev. \textbf{D\,19}, 3293 (1979)

\bibitem{bib:lahmann-habil-2011}
R.~Lahmann, \emph{{Ultra-High-Energy Neutrinos and Their Acoustic Detection in
  the Sea}}, Habilitation thesis (2011)

\bibitem{bib:baikal-arena2010}
{V.~Aynutdinov~\etal}, Nucl.\ Inst.\ and Meth.\ \textbf{A\ 662}, S210 (2012),
  doi:10.1016/j.nima.2010.11.153

\bibitem{bib:Liebermann-1949}
{L.~Liebermann}, Phys.\ Rev. \textbf{76}, 1520 (1949)

\bibitem{urick}
{R.J.\ Urick}, \emph{Principles of Underwater Sound} (Peninsula publishing,
  1983), {ISBN 0-932146-62-7}

\bibitem{bib:Ainslie-98}
{M.A.~Ainslie and J.G.~McColm}, J.\ Acoust.\ Soc.\ Am. \textbf{103}, 1671
  (1998)

\bibitem{anton}
{G.~Anton \etal}, Astropart.\ Phys. \textbf{26}, 301 (2006)

\bibitem{bib:spats-2011-bkgr}
{R.~Abbasi \etal\ (IceCube Coll.)}, Astropart.\ Phys. \textbf{35}, 312 (2012)

\bibitem{bib:saund2010}
{N.~Kurahashi, J.~Vandenbroucke, and G.~Gratta}, Phys.\ Rev. \textbf{D\,82},
  073006 (2010)

\bibitem{bib:bevan_ARENA08}
{S.~Bevan}, Nucl.\ Inst.\ and Meth.\ \textbf{A\,604}, 143 (2009)

\bibitem{bib:karg-arena2010}
{T.~Karg, for the IceCube Coll.}, {Nucl.\ Inst.\ and Meth.\ } \textbf{A\,662},
  S36 (2012)

\bibitem{bib:SPATS-design-perform-2011}
{Y.~Abdou~\etal\ (IceCube Coll.)}, Nucl.\ Inst.\ and Meth.\ A \textbf{683}, 78
  (2012)

\bibitem{bib:noise_ONDE_arxiv}
{S.~Aiello~\etal (NEMO Coll.), L.~Cosentino~\etal} (2008), {arXiv:0804.2913
  [astro-ph]}

\bibitem{bib:amadeus-2010}
{J.A. Aguilar~\etal\ (ANTARES Coll.)}, Nucl.\ Inst.\ and Meth. \textbf{A
  626-627}, 128 (2011)

\bibitem{bib:acorne}
{S.~Danaher for the ACoRNE Coll. }, J.\ Phys.\ Conf.\ Ser. \textbf{81}, 012011
  (2007)

\bibitem{bib:Lehtinen-2002}
{N.G. Lehtinen \etal}, Astropart.\ Phys. \textbf{17}, 279 (2002),
  {arXiv:astro-ph/0104033v1}

\bibitem{bib:km3net_acou_arena2016}
{F.\ Simeone}, \emph{{Acoustic detection of UHE neutrinos in the Mediterranean
  sea: status and perspective}}, in \emph{{Proc.\ of ARENA 2016 workshop}}
  (2016), to be published in EPJ Web of Conf.

\bibitem{bib:KM3NeT2p0_LOI_2016}
{S.~Adri\'an-Mar\'inez \etal (KM3NeT Coll.)}, J.\ Phys.\ G: Nucl.\ Part.\ Phys.
  \textbf{43}, 084001 (2016)

\bibitem{bib:buis_arena2016}
{E.J.~Buis}, \emph{{A large fiber sensor network for an acoustic neutrino
  telescope}}, in \emph{{Proc.\ of ARENA 2016 workshop}} (2016), to be
  published in EPJ Web of Conf.

\bibitem{urick2}
{R.J.\ Urick}, \emph{Ambient Noise in the Sea} (Peninsula publishing, 1986),
  {ISBN 0-932146-13-9}

\bibitem{Kiessling_master-2013}
D.~Kie{\ss}ling, Master's thesis, Univ.\ Erlangen-N{\"u}rnberg (2013),
  {ECAP-2013-041}

\bibitem{bib:lahmann-amadeus-icrc2011}
{R.~Lahmann for the ANTARES Coll.} (2011), arXiv:1104.3041 [astro-ph.IM]

\bibitem{bib:Neff_PhD_2013}
{M.~Neff}, Ph.D. thesis, Univ.\ Erlangen-N{\"u}rnberg (2013), {ECAP-2013-023}

\bibitem{bib:kiessling_arena2016}
{D.\ Kie\ss ling}, \emph{{Signal classification and event reconstr.\ for
  acoustic neutrino detection in sea water with KM3NeT}}, in \emph{{Proc.\ of
  ARENA 2016 workshop}} (2016), to be published in EPJ Web of Conf.

\end{thebibliography}


\begin{thebibliography}{}
%
%
\bibitem{RefJ}
Journal Author, Journal \textbf{Volume}, page numbers (year)
\bibitem{RefB}
Book Author, \textit{Book title} (Publisher, place, year) page numbers
\end{thebibliography}
\comment{
%

} 

\end{document}